\documentclass[prl,twocolumn,showpacs,amsmath,amssymb,superscriptaddress]{revtex4}
\usepackage{pifont,amssymb,epsf,graphicx}
\begin{document}
\title{Model for Tunneling-mediated Impurity Resonances in Bilayer Cuprate Superconductors}
\author{Degang Zhang}
\affiliation{Texas Center for Superconductivity and Department 
of Physics, University of Houston, Houston, TX 77204, USA}
\author{C. S. Ting}
\affiliation{Texas Center for Superconductivity and Department 
of Physics, University of Houston, Houston, TX 77204, USA}

%\date{\today}
\begin{abstract}

We have studied tunneling-mediated local density of states (LDOS) of the surface layer of a bilayer cuprate, where a Zn impurity is located on the second Cu-O layer. When the tunneling strength between two Cu-O layers is larger than a critical value, the LDOS on the site just above the Zn impurity first exhibits a resonant peak near the Fermi surface.  The larger the tunneling strength, the stronger the resonant peak. It is also shown that the height of the resonant peak oscillates decreasingly with the distance from the site just above the Zn impurity. The location of the resonant peak in the surface LDOS depends on doping, energy gap, and the tunneling strength, and has an opposite bias voltage to that on its nearest neighboring sites. The results could be tested by the STM experiments and be used to further understand the electronic properties of high temperature superconductors. 

\end{abstract}

\maketitle

High temperature superconductivity in cuprates has been the focus of theoretical and experimental investigations [1-7]. One of the main features of the cuprate superconductors is that their electronic properties depend strongly on the number of Cu-O planes in a unit cell. The bilayer LDA band calculations predicted that the c-axis electron hopping has an anisotropic form $t_{\bf k}=-t_z(\cos {k}_{x}-\cos {k}_{y})^2/4$, where $t_z$ is the tunneling strength [8]. Recently angle-resolved photoemission spectroscopy (ARPES) [2], one direct probe of the electronic structure in the momentum space, confirmed the bilayer splitting in the electronic structure of (heavily) overdoped Bi$_2$ Sr$_2$CaCu$_2$O$_{8+\delta}$ (Bi2212) [9,10]. The experimental value of $t_z$ is of similar magnitude of the superconducting energy gap. 

However, scanning tunneling microscopy (STM) [7], one direct probe of local density of states (LDOS) at the surface of samples in the real space, has not yet reported the distinct feature in bilayer cuprates. This is because it is very difficult to figure out the bilayer splitting in uniform samples by the STM experiments. In Ref. [11], Pan {\it et al.} observed a resonant peak near the Fermi surface on the Zn impurity located at the first Cu-O layer of Bi2212 and discovered a four-fold symmetric LDOS pattern by the STM experiment. The LDOS pattern near the Zn impurity is consistent with the symmetry of the superconducting order parameter. These experimental phenomena were explained by a unitary potential and the blocking effect due to the Bi atom just above the Zn impurity [12] (For an alternative model, see Ref. [13]). In this work, we investigate the surface LDOS in bilayer Bi2212, where a Zn impurity is located on the second Cu-O layer so that we can find out the features of the tunneling-mediated impurity states and the tunneling splitting, which could be observed by the STM experiments.

The Hamiltonian describing the scattering of quasiparticles from a single impurity and the inter-layer tunneling in a d-wave bilayer superconductor can be written as
$$H=\sum_{{u,\bf k}\sigma }{\epsilon}_{\bf k}
{c}_{u,{\bf k}\sigma }^{+}{c}_{u,{\bf k}\sigma
}+\sum_{u,{\bf k}}{\Delta }_{{\bf k}}(c_{u,{\bf k}
\uparrow }^{+}c_{u,-{\bf k}\downarrow }^{+}+c_{u,-{\bf k}
\downarrow }c_{{u,\bf k}\uparrow })$$
$$+\sum_{{\bf k},\sigma }t_{\bf k}({c}_{0,{\bf k}\sigma }^{+}{c}_{1,{\bf k}\sigma}
+{c}_{1,{\bf k}\sigma }^{+}{c}_{0,{\bf k}\sigma})
 +V_{s}\sum_{\sigma}c_{1,0\sigma }^{+}c
_{1,0\sigma},\eqno{(1)}$$
where $u=0$ and 1 represent the first and second Cu-O layers, respectively, $\sigma$ is the spin index of electrons, $c_{u,{\bf r}\sigma}=\frac{1}{\sqrt{N}}\sum_{\bf k}c_{u,{\bf k}\sigma}e^{i{\bf k}\cdot{\bf r}}$, $N$ is the number of sites in the lattice and is taken as $400\times 400$ in our calculations below,
$\epsilon_{{\bf k}}=t_{1}(\cos {k}_{x}+\cos {k}_{y})/2+t_{2}\cos 
{k}_{x}\cos {k}_{y}+t_{3}(\cos 2{k}_{x}+\cos 2
{k}_{y})/2+t_{4}(\cos 2{k}_{x}\cos {k}_{y}+\cos 
{k}_{x}\cos 2{k}_{y})/2+t_{5}\cos 2{k}_{x} \cos 2
{k}_{y}-\mu $ with $t_{1-5}$=-0.5951, 0.1636, -0.0519, -0.1117, 0.0510 (eV). These band parameters are taken from APRES measurement for Bi2212 [14], and the lattice constant $a$ is set as $a$=1. $\mu $ is the chemical potential to be determined by doping $p$. The superconducting order parameter $\Delta _{{\bf k}}=\Delta _{0}(\cos {k}_{x}-\cos {k}_{y})/2 $. 
$V_s$ is the strength of nonmagnetic potential.
 
In order to solve the Hamiltonian (1), we first take the Bogoliubov transformation: 
$c_{u,{\bf k}\uparrow }=\sum_{\nu=0,1}(-1)^{\nu}\xi_{{\bf k}\nu}\psi_{u,{\bf k}\nu},
c_{u,-{\bf k}\downarrow }^{+}=\sum_{\nu=0,1}\xi_{{\bf k}\nu}\psi_{u,{\bf k}\nu+1},$
where 
$\xi^2_{{\bf k}\nu}=\frac{1}{2}[1+(-1)^\nu\frac{{\epsilon}_{\bf k}}{E_{\bf k}}]$,
$\xi_{{\bf k}\nu}\xi_{{\bf k}\nu+1}=\frac{\Delta_{\bf k}}{2E_{\bf k}}$,
and $E_{\bf k}=\sqrt{{\epsilon}_{\bf k}^2+\Delta ^2_{{\bf k}}}$,
then the Hamiltonian (1) can be re-expressed in terms of the fermionic creation and annihilation operators $\psi^{+}_{u,{\bf k}\nu}$ and $\psi_{u,{\bf k}\nu}$.

We define two-point Green's functions
$$G^{u^\prime,{\bf k}^\prime\nu^\prime}_{u,{\bf k}\nu}(i\omega_n)=-{\cal F}<T_\tau\psi_{u,{\bf k}\nu}(\tau)\psi_{u^\prime,{\bf k}^\prime\nu^\prime}^+(0)>,\eqno{(2)}$$
where ${\cal F}\phi(\tau)$ denote the Fourier transform of $\phi(\tau)$ in Matsubara frequencies, $\tau$ is the imaginary time. Following the standard approach [15], the Green's functions $G^{u^\prime,{\bf k}^\prime\nu^\prime}_{u,{\bf k}\nu}(i\omega_n)$ satisfy  

$$[G^0_{{\bf k}\nu}]^{-1}G^{u^\prime,{\bf k}^\prime\nu^\prime}_{u,{\bf k}\nu}-\frac{t_{\bf k}}{E_{\bf k}}[(-1)^{\nu}\epsilon_{{\bf k}}G^{u^\prime,{\bf k}^\prime\nu^\prime}_{u+1,{\bf k}\nu}-\Delta_{{\bf k}}G^{u^\prime,{\bf k}^\prime \nu^\prime}_{u+1,{\bf k}\nu+1}]  $$

$$-\frac{V_s}{N}\delta_{u1}\sum_{{\bf k}^{\prime\prime}\nu^{\prime\prime}}[(-1)^{\nu+\nu^{\prime\prime}}\xi_{{\bf k}\nu}\xi_{{\bf k}^{\prime\prime}\nu^{\prime\prime}}
-\xi_{{\bf k}\nu+1}\xi_{{\bf k}^{\prime\prime}\nu^{\prime\prime}+1}]G^{{u^\prime,\bf k}^\prime\nu^\prime}_{u,{\bf k}^{\prime\prime}\nu^{\prime\prime}}$$
$$=\delta_{uu^\prime}\delta_{{\bf k}{\bf k}^\prime}\delta_{\nu\nu^\prime},
\eqno{(3)}$$
where the bare Green's functions $G^0_{{\bf k}\nu}=[i\omega_n-(-1)^{\nu}E_{\bf k}]^{-1}$. From Eq. (3), we have the Green's functions with $u=0$

$$G^{u^\prime,{\bf k}^\prime\nu^\prime}_{0,{\bf k}\nu}=G^0_{{\bf k}\nu}\delta_{0u^\prime}\delta_{{\bf k}{\bf k}^\prime}\delta_{\nu\nu^\prime}$$
$$+\frac{t_{\bf k}}{E_{\bf k}}G^0_{{\bf k}\nu}[(-1)^{\nu}\epsilon_{{\bf k}}G^{u^\prime,{\bf k}^\prime\nu^\prime}_{1,{\bf k}\nu}-\Delta_{{\bf k}}G^{u^\prime,{\bf k}^\prime \nu^\prime}_{1,{\bf k}\nu+1}],
\eqno{(4)}$$
which are related to the Green's functions with $u=1$. Obviously, when the tunneling strength $t_z=0$, $G^{u^\prime,{\bf k}^\prime\nu^\prime}_{0,{\bf k}\nu}$ become the bare Green's functions $G^0_{{\bf k}\nu}$. Therefore, any impurity (mid-gap) state in the first Cu-O layer is produced through the inter-layer tunneling. In other words, the existence of resonant peak in the surface LDOS provides a direct evidence of the tunneling splitting in bilayer cuprates. Substituting the expression of $G^{{u^\prime,\bf k}^\prime\nu^\prime}_{0,{\bf k}\nu}(i\omega_n)$ into Eq. (3) with $u=1$, we have two linearly dependent equations in terms of $G^{{u^\prime,\bf k}^\prime\nu^\prime}_{1,{\bf k}0}$ and $G^{{u^\prime,\bf k}^\prime\nu^\prime}_{1,{\bf k}1}$, i.e. 

$$a_{{\bf k}\nu}G^{{u^\prime,\bf k}^\prime\nu^\prime}_{1,{\bf k}\nu}
+b_{{\bf k}\nu}G^{{u^\prime,\bf k}^\prime\nu^\prime}_{1,{\bf k}\nu+1}
-(-1)^\nu \frac{V_s}{N}\xi_{{\bf k}\nu}G^0_{{\bf k}\nu}\alpha_{u^\prime,{\bf k}^\prime\nu^\prime}$$

$$+\frac{V_s}{N}\xi_{{\bf k}\nu+1}G^0_{{\bf k}\nu}\beta_{u^\prime,{\bf k}^\prime\nu^\prime}
=\lambda^{u^\prime,{\bf k}^\prime\nu^\prime}_{\Box,{\bf k}\nu},
\eqno{(5)}$$
where we have set

$$ a_{{\bf k}\nu}=1-(\frac{t_{\bf k}}{E_{\bf k}})^2G^0_{{\bf k}\nu}(\epsilon_{{\bf k}}^2G^0_{{\bf k}\nu}
  +\Delta_{{\bf k}}^2G^0_{{\bf k}\nu+1}),$$

$$ b_{{\bf k}\nu}=(-1)^{\nu}(\frac{t_{\bf k}}{E_{\bf k}})^2\epsilon_{{\bf k}}\Delta_{{\bf k}}G^0_{{\bf k}\nu}(G^0_{{\bf k}\nu}
  -G^0_{{\bf k}\nu+1}),$$

$$\alpha_{u^\prime,{\bf k}^\prime\nu^\prime}=\sum_{{\bf k}^{\prime\prime}\nu^{\prime\prime}}(-1)^{\nu^{\prime\prime}}\xi_{{\bf k}^{\prime\prime}\nu^{\prime\prime}}
G^{{u^\prime,\bf k}^\prime\nu^\prime}_{1,{\bf k}^{\prime\prime}\nu^{\prime\prime}}, $$

$$\beta_{u^\prime,{\bf k}^\prime\nu^\prime}=\sum_{{\bf k}^{\prime\prime}\nu^{\prime\prime}}\xi_{{\bf k}^{\prime\prime}\nu^{\prime\prime}+1}
G^{{u^\prime,\bf k}^\prime\nu^\prime}_{1,{\bf k}^{\prime\prime}\nu^{\prime\prime}}, $$

$$\lambda^{u^\prime,{\bf k}^\prime\nu^\prime}_{\Box,{\bf k}\nu}=G^0_{{\bf k}\nu}\delta_{{\bf k}{\bf k}^\prime}\{\delta_{1u^\prime}\delta_{\nu\nu^\prime}$$
$$+\frac{t_{\bf k}}{E_{\bf k}}\delta_{0u^\prime}[(-1)^{\nu}\epsilon_{{\bf k}}G^0_{{\bf k}\nu}\delta_{\nu\nu^\prime}
 -\Delta_{{\bf k}}G^0_{{\bf k}\nu+1}\delta_{\nu+1\nu^\prime}]\}.
\eqno{(6)}$$
Solving Eq. (5), we obtain the Green's functions with $u=1$

$$G^{u^\prime,{\bf k}^\prime\nu^\prime}_{1,{\bf k}\nu}=[
(-1)^\nu \frac{V_s}{N}(\xi_{{\bf k}\nu}a_{{\bf k}\nu+1}G^0_{{\bf k}\nu}+\xi_{{\bf k}\nu+1}b_{{\bf k}\nu}G^0_{{\bf k}\nu+1})\alpha_{u^\prime,{\bf k}^\prime\nu^\prime}$$
$$-\frac{V_s}{N}(\xi_{{\bf k}\nu+1}a_{{\bf k}\nu+1}G^0_{{\bf k}\nu}-\xi_{{\bf k}\nu}b_{{\bf k}\nu}G^0_{{\bf k}\nu+1})\beta_{u^\prime,{\bf k}^\prime\nu^\prime}
+\Gamma^{u^\prime,{\bf k}^\prime\nu^\prime}_{\Box,{\bf k}\nu}]D_{{\bf k}\nu},
\eqno{(7)}$$
where $D_{{\bf k}\nu}=(a_{{\bf k}\nu}a_{{\bf k}\nu+1}-b_{{\bf k}\nu}b_{{\bf k}\nu+1})^{-1}$ and $\Gamma^{u^\prime,{\bf k}^\prime\nu^\prime}_{\Box,{\bf k}\nu}=a_{{\bf k}\nu+1}\lambda^{u^\prime,{\bf k}^\prime\nu^\prime}_{\Box,{\bf k}\nu}-b_{{\bf k}\nu}\lambda^{u^\prime,{\bf k}^\prime\nu^\prime}_{\Box,{\bf k}\nu+1}$. 
We note that $\alpha_{u^\prime,{\bf k}^\prime\nu^\prime}$ and $\beta_{u^\prime,{\bf k}^\prime\nu^\prime}$ in the expression of $G^{u^\prime,{\bf k}^\prime\nu^\prime}_{1,{\bf k}\nu}$ are undetermined.
Multiplying Eq. (7) by $(-1)^\nu\xi_{{\bf k}\nu}$ and $\xi_{{\bf k}\nu+1}$, respectively, and then summating ${\bf k}$ and $\nu$, we also have two linearly dependent equations in terms of 
$\alpha_{u^\prime,{\bf k}^\prime\nu^\prime}(i\omega_n)$ and $\beta_{u^\prime,{\bf k}^\prime\nu^\prime}(i\omega_n)$, which have solutions

$$ \alpha_{u^\prime,{\bf k}^\prime\nu^\prime}=C(c_{22}d^1_{u^\prime,{\bf k}^\prime\nu^\prime}-c_{12}d^2_{u^\prime,{\bf k}^\prime\nu^\prime}),$$
$$ \beta_{u^\prime,{\bf k}^\prime\nu^\prime}=C(c_{11}d^2_{u^\prime,{\bf k}^\prime\nu^\prime}-c_{21}d^1_{u^\prime,{\bf k}^\prime\nu^\prime}).
\eqno{(8)}$$
Here we have defined
$$c_{11}=1-\frac{V_s}{N}\sum_{{\bf k}\nu}D_{{\bf k}\nu}\xi_{{\bf k}\nu}(\xi_{{\bf k}\nu}a_{{\bf k}\nu+1}G^0_{{\bf k}\nu}+\xi_{{\bf k}\nu+1}b_{{\bf k}\nu}G^0_{{\bf k}\nu+1}),$$
$$c_{12}=\frac{V_s}{N}\sum_{{\bf k}\nu}(-1)^{\nu}D_{{\bf k}\nu}\xi_{{\bf k}\nu}(\xi_{{\bf k}\nu+1}a_{{\bf k}\nu+1}G^0_{{\bf k}\nu}-\xi_{{\bf k}\nu}b_{{\bf k}\nu}G^0_{{\bf k}\nu+1}),$$
$$c_{21}=-\frac{V_s}{N}\sum_{{\bf k}\nu}(-1)^{\nu}D_{{\bf k}\nu}\xi_{{\bf k}\nu+1}(\xi_{{\bf k}\nu}a_{{\bf k}\nu+1}G^0_{{\bf k}\nu}+\xi_{{\bf k}\nu+1}b_{{\bf k}\nu}G^0_{{\bf k}\nu+1}),$$
$$c_{22}=1+\frac{V_s}{N}\sum_{{\bf k}\nu}D_{{\bf k}\nu}\xi_{{\bf k}\nu+1}[\xi_{{\bf k}\nu+1}a_{{\bf k}\nu+1}G^0_{{\bf k}\nu}-\xi_{{\bf k}\nu}b_{{\bf k}\nu}G^0_{{\bf k}\nu+1}],$$
$$C=(c_{11}c_{22}-c_{12}c_{21})^{-1},$$
$$d^1_{u^\prime,{\bf k}^\prime\nu^\prime}=\sum_{{\bf k}\nu}(-1)^{\nu}D_{{\bf k}\nu}\xi_{{\bf k}\nu}\Gamma^{u^\prime,{\bf k}^\prime\nu^\prime}_{\Box,{\bf k}\nu}, $$
$$d^2_{u^\prime,{\bf k}^\prime\nu^\prime}=\sum_{{\bf k}\nu}D_{{\bf k}\nu}\xi_{{\bf k}\nu+1}\Gamma^{u^\prime,{\bf k}^\prime\nu^\prime}_{\Box,{\bf k}\nu}. 
\eqno{(9)}$$
Up to now, we have obtained the Green's functions $G^{u^\prime,{\bf k}^\prime\nu^\prime}_{u,{\bf k}\nu}(\pm i\omega_n)$. Obviously, the Green's functions have the symmetries $G^{u^\prime,{\bf k}^\prime\nu^\prime}_{u,{\bf k}\nu}(\pm i\omega_n)\equiv G^{u^\prime,{\bf k}\nu}_{u,{\bf k}^\prime\nu^\prime}(\pm i\omega_n)$, which enable us to simplify the calculation of the LDOS below. 

\begin{figure}
\rotatebox[origin=c]{0}{\includegraphics[angle=0, 
           height=1.3in]{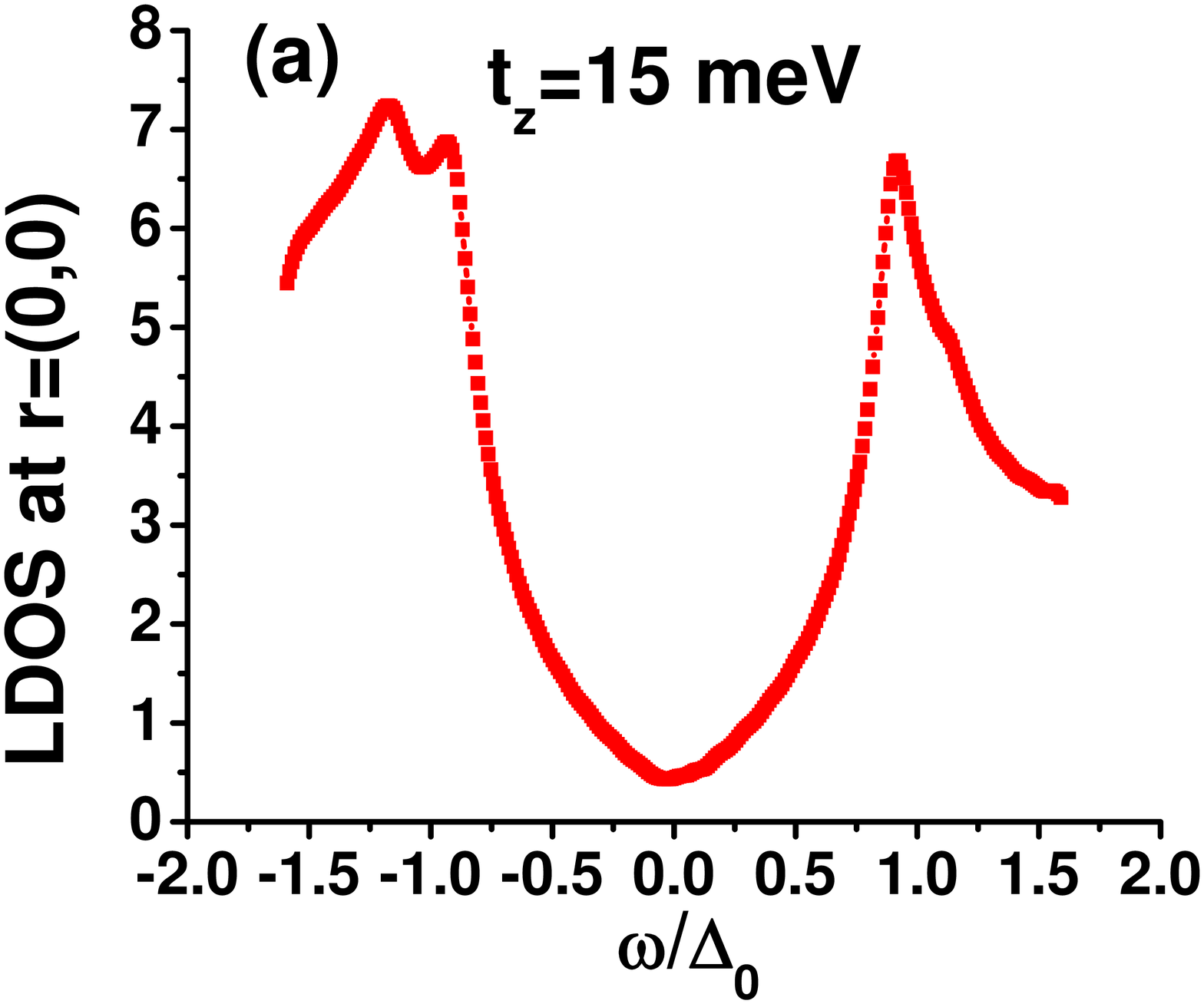}}
\rotatebox[origin=c]{0}{\includegraphics[angle=0, 
           height=1.3in]{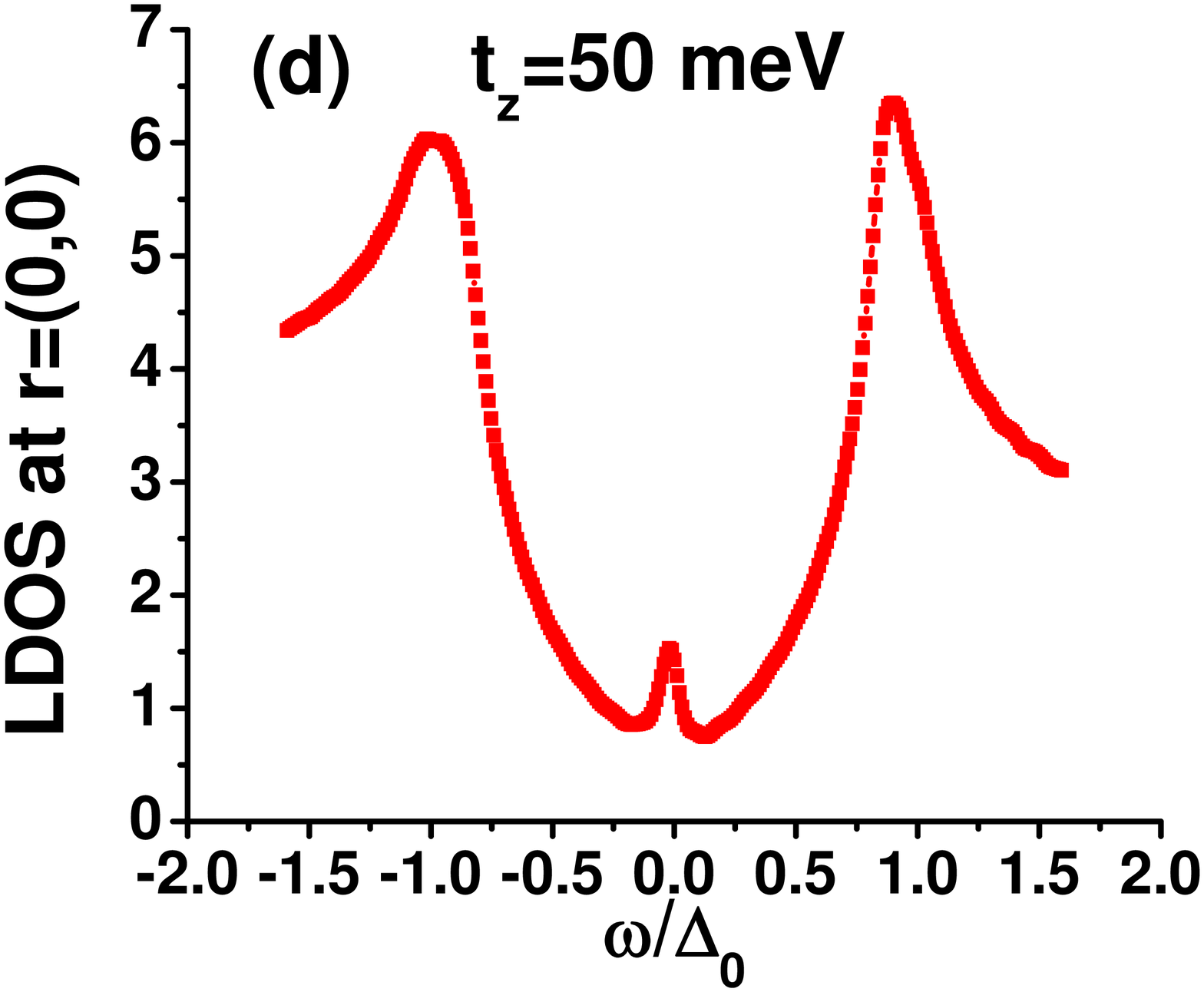}}
\rotatebox[origin=c]{0}{\includegraphics[angle=0, 
           height=1.3in]{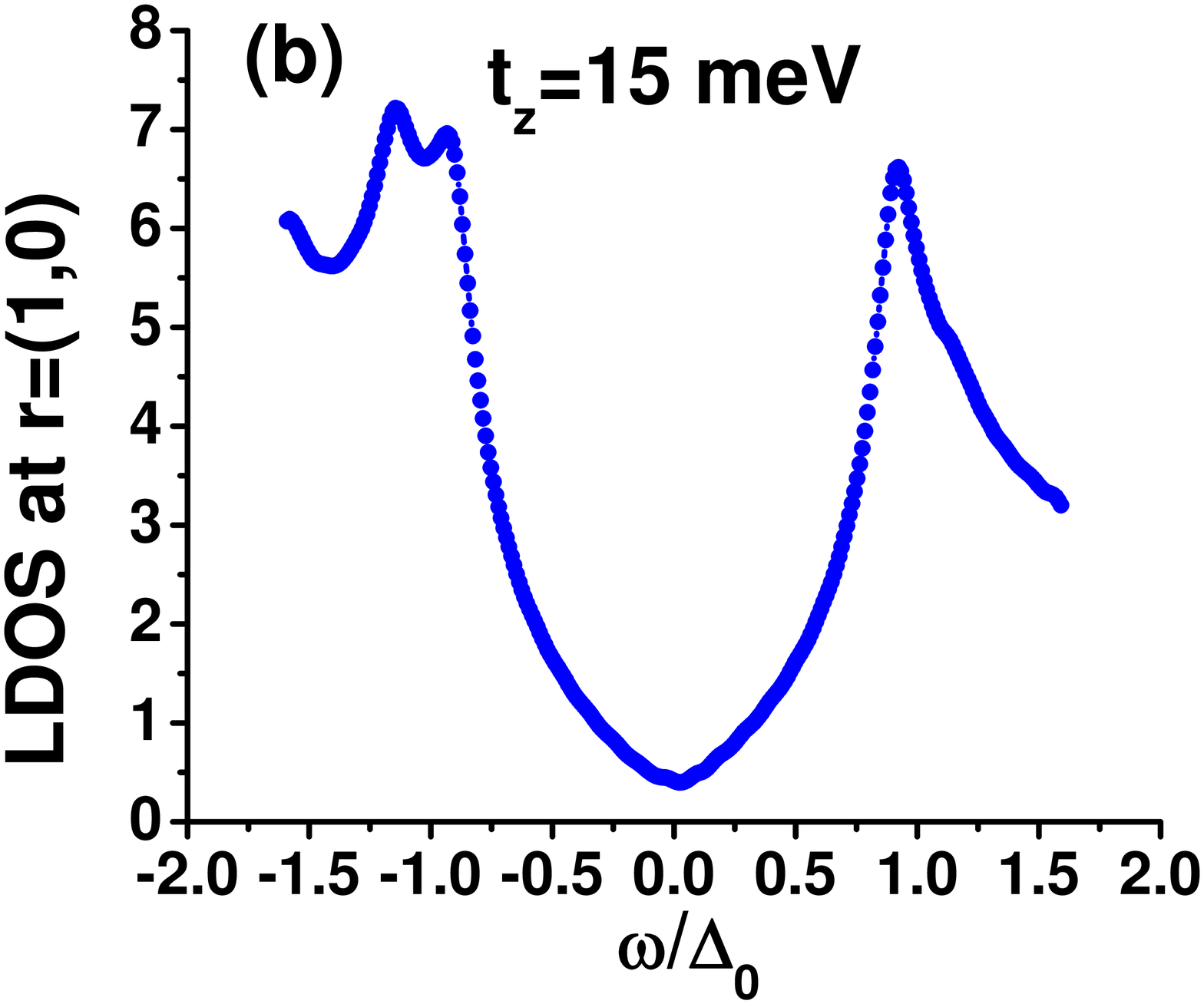}}
\rotatebox[origin=c]{0}{\includegraphics[angle=0, 
           height=1.3in]{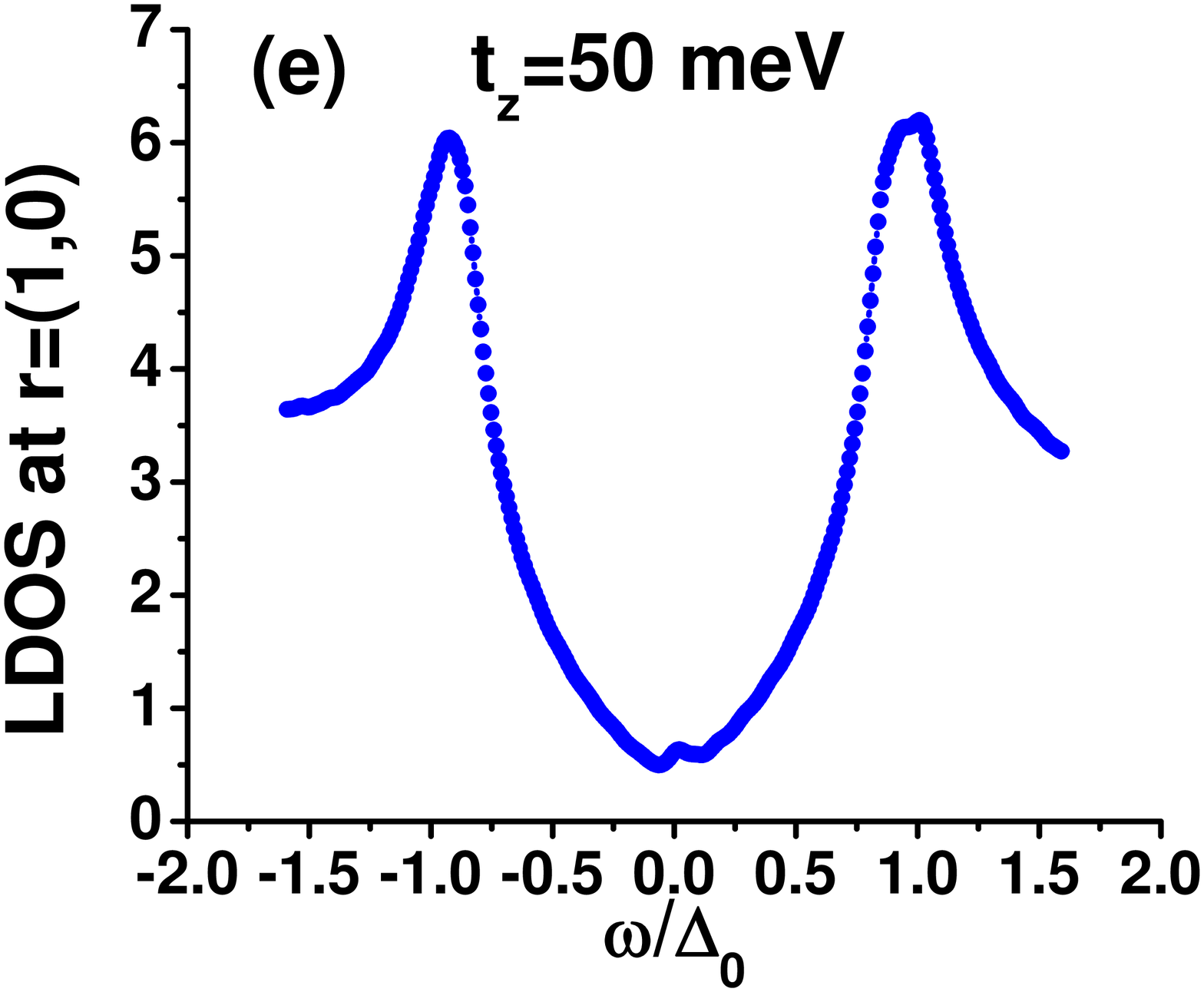}}
\rotatebox[origin=c]{0}{\includegraphics[angle=0, 
           height=1.3in]{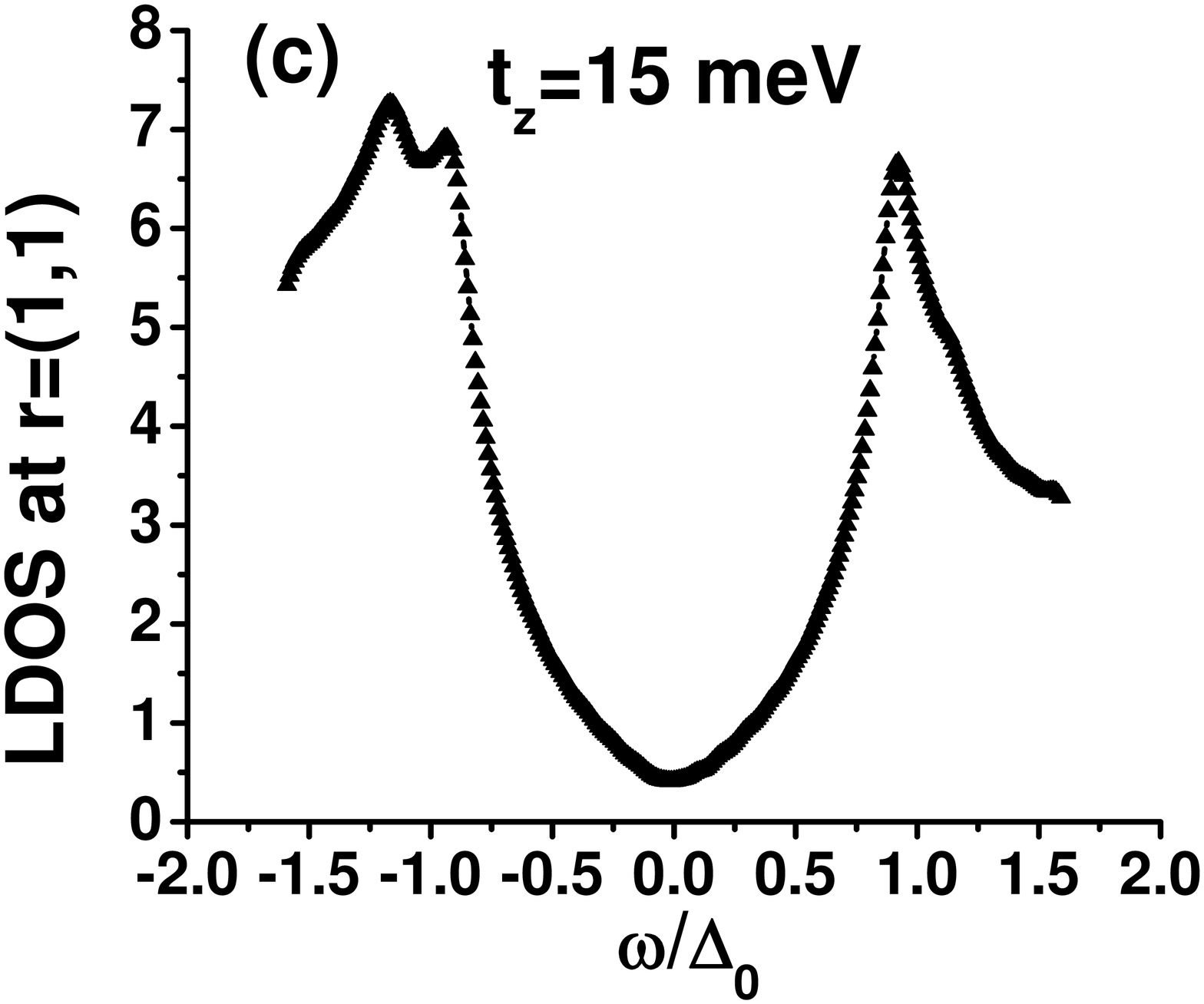}} 
\rotatebox[origin=c]{0}{\includegraphics[angle=0, 
           height=1.3in]{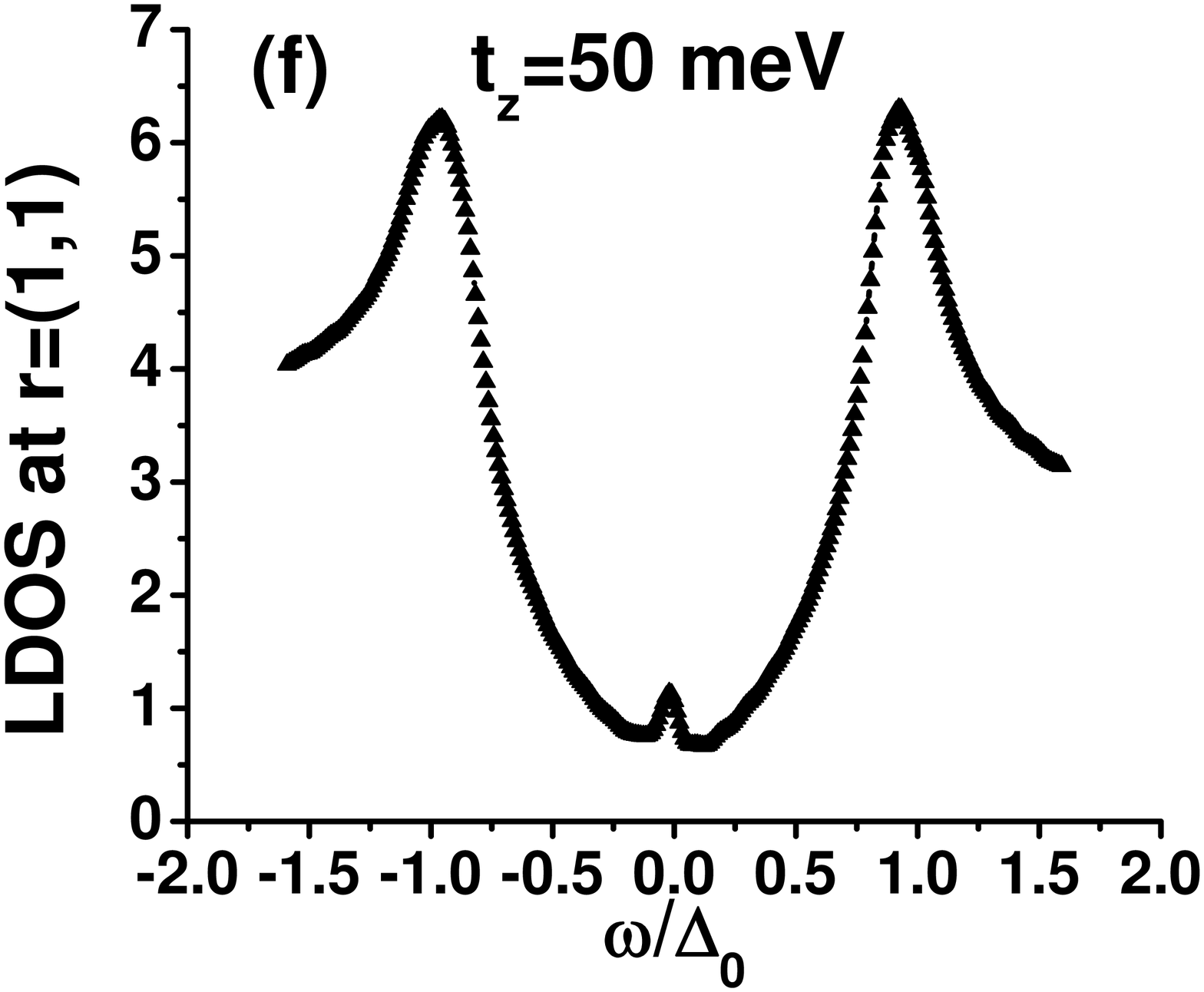}}              
\caption{(Color online) The LDOS $\rho({\bf r},\omega)$ at the sites (0,0), (1,0), and (1,1) with different tunneling strength $t_z$. The energy gap $\Delta_0=44$ meV and the doping $p\sim 15\%$.}           
\end{figure}

\begin{figure}
\rotatebox[origin=c]{0}{\includegraphics[angle=0, 
           height=2.6in]{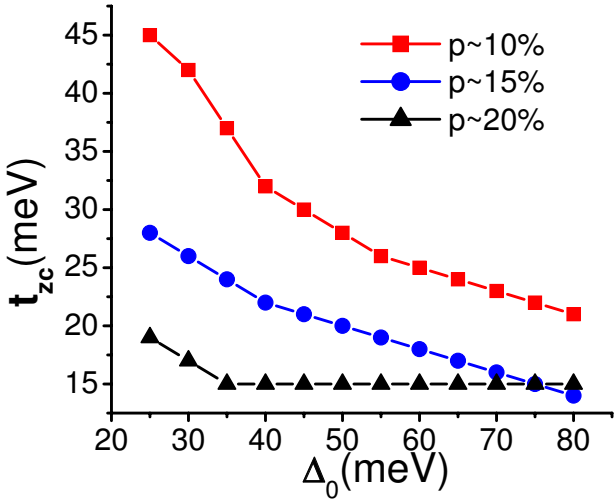}}                   
\caption {(Color online) The critical value $t_{zc}$ of the tunneling strength as functions of the energy gap $\Delta_0$ and the doping $p$.}
\end{figure}

The LDOS at the first Cu-O layer (i.e. $u=0$) can be expressed as [16,17]
$$\rho({\bf r},\omega)=-\frac{1}{\pi}{\rm Im} \sum_{\sigma}[-{\cal F}<c_{0,{\bf r}\sigma}(\tau)c_{0,{\bf r}\sigma}^+(0)>]|_{i\omega_n\rightarrow \omega+i0^+}$$
$$=-\frac{1}{N\pi}{\rm Im} \sum_{{\bf k},{\bf k}^\prime,\nu,\nu^\prime}\cos [({\bf k}-{\bf k}^\prime)
\cdot {\bf r}][(-1)^{\nu+\nu^\prime}\xi_{{\bf k}\nu}\xi_{{\bf k}^\prime\nu^\prime}$$
$$\times G^{0,{\bf k}^\prime\nu^\prime}_{0,{\bf k}\nu}(i\omega_n)-\xi_{{\bf k}\nu+1}\xi_{{\bf k}^\prime\nu^\prime+1}G^{0,{\bf k}^\prime\nu^\prime}_{0,{\bf k}\nu}(-i\omega_n)]|_{i\omega_n\rightarrow \omega+i0^+}.\eqno{(10)}$$
Combining Eqs. (4), (7) and (10), we can calculate the curves of the LDOS with various physical parameters at different sites and images of the LDOS at different bias voltages. In Fig. 1, we present $\rho({\bf r},\omega)$ at the sites (0,0), (1,0), and (1,1) when the tunneling strength $t_z=15$ meV and 50 meV, respectively. When $t_z=15$ meV, $\rho({\bf r},\omega)$ at these sites are identical below energy gap. When $t_z=50$ meV, $\rho({\bf r},\omega)$ has a resonant peak at (0,0) and (1,1) with $\omega=-0.8$ meV and at (1,0) with $\omega=0.8$ meV. The height of the resonant peaks oscillates decreasingly with the distance from the site (0,0) just above the Zn impurity. We note that these resonant peaks become higher with increasing $t_z$. Undoubtedly, these impurity resonances are unique features of the tunneling between two Cu-O layers.  

\begin{figure}
\rotatebox[origin=c]{0}{\includegraphics[angle=0, 
           height=1.3in]{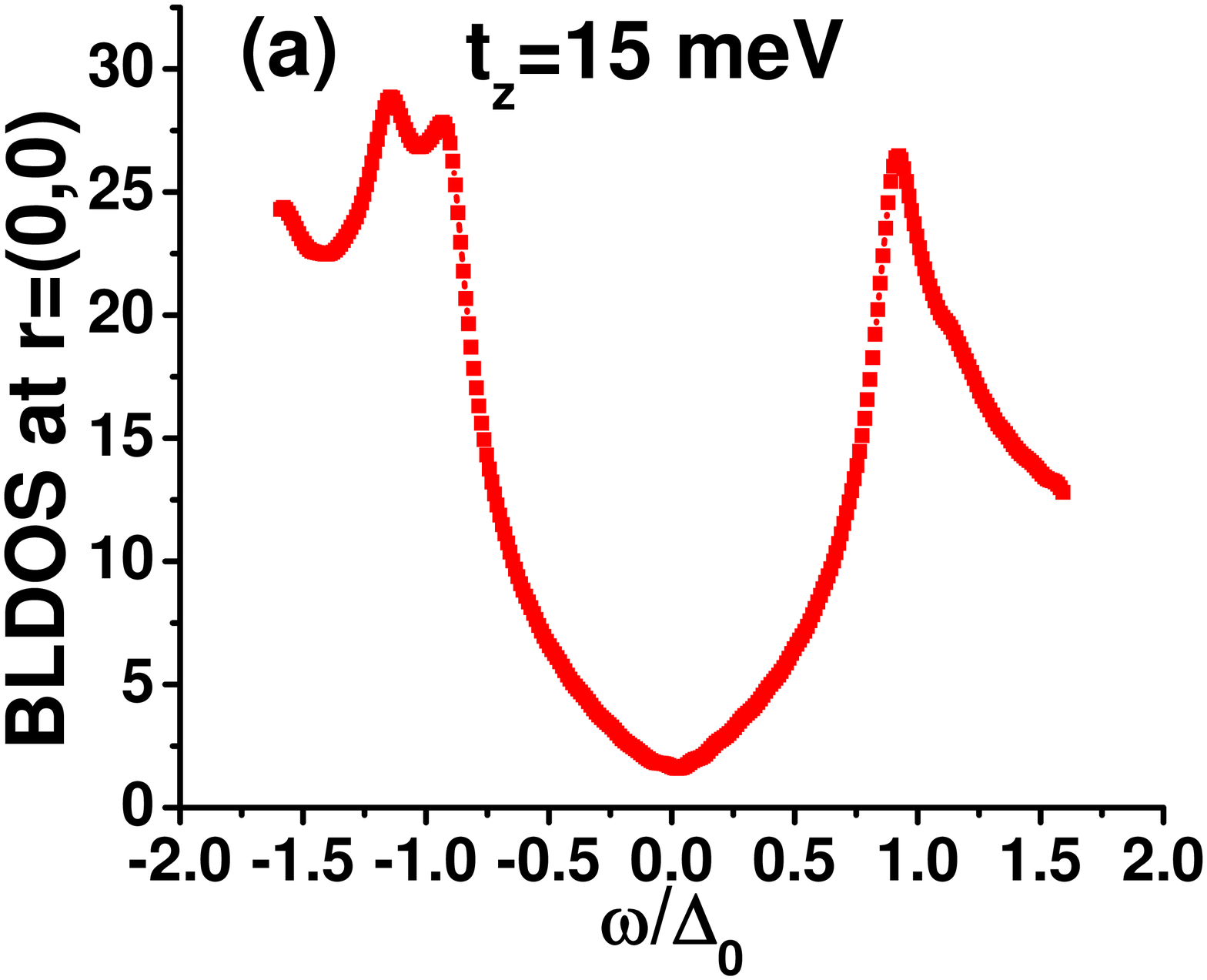}}
\rotatebox[origin=c]{0}{\includegraphics[angle=0, 
           height=1.3in]{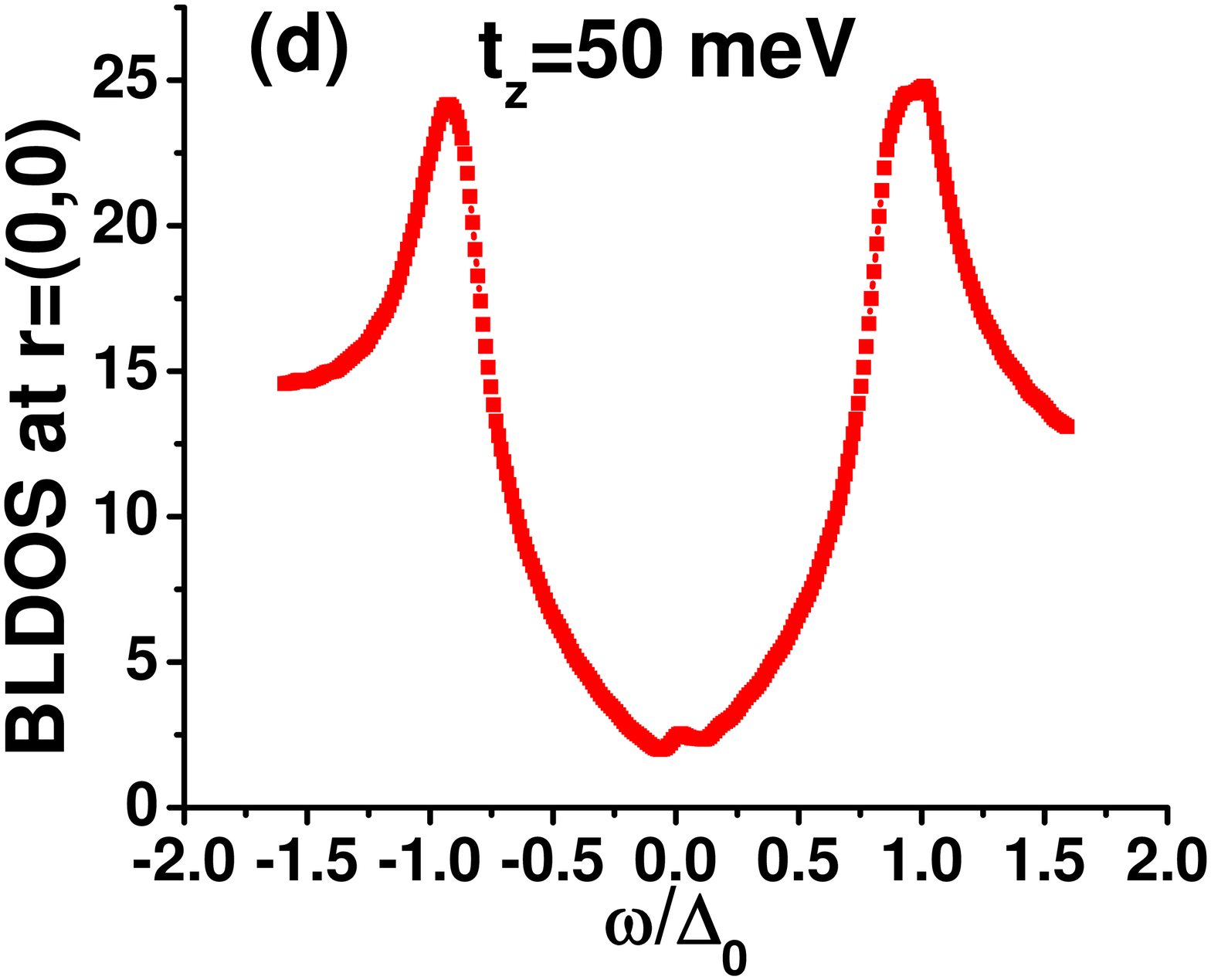}}
\rotatebox[origin=c]{0}{\includegraphics[angle=0, 
           height=1.3in]{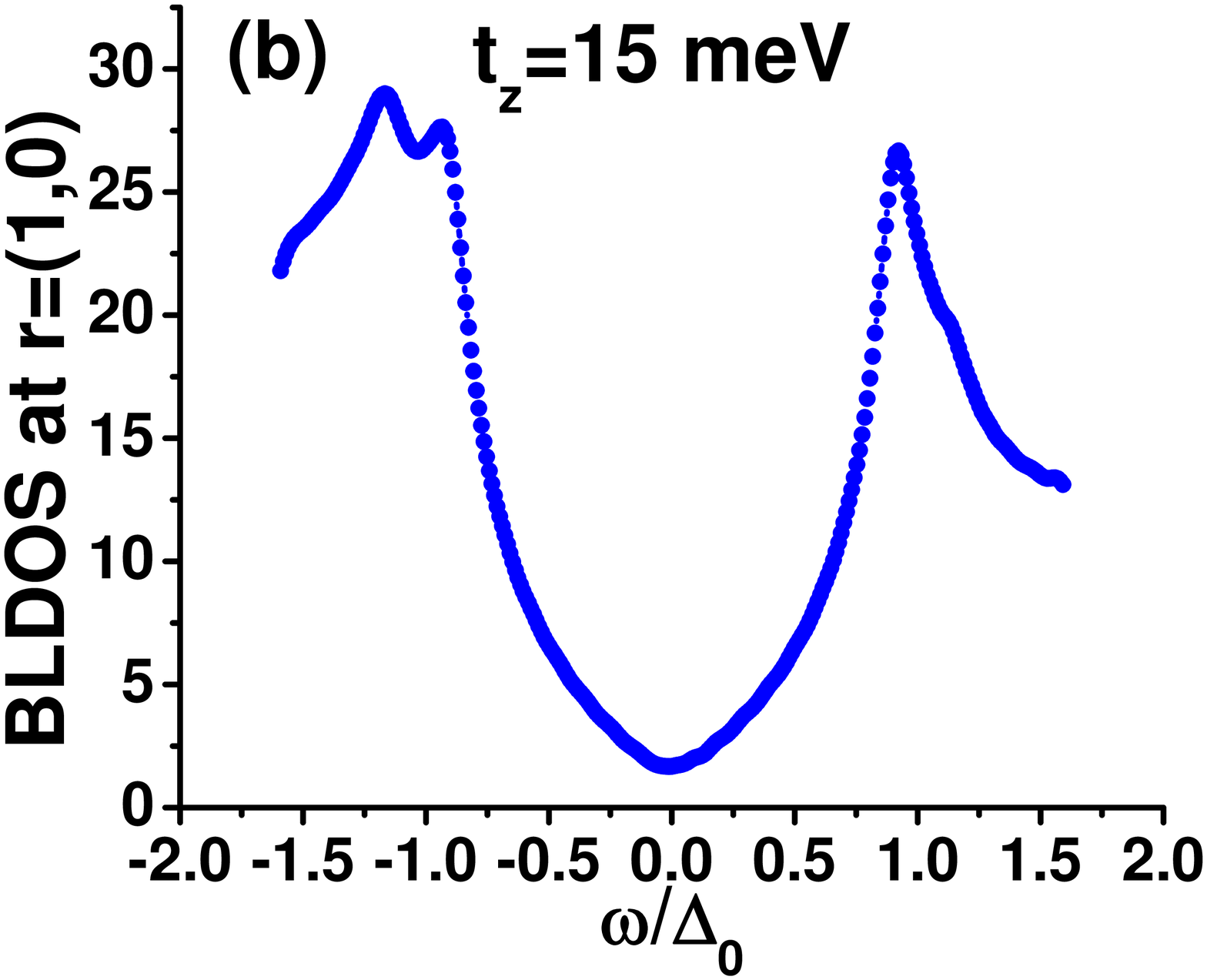}}
\rotatebox[origin=c]{0}{\includegraphics[angle=0, 
           height=1.3in]{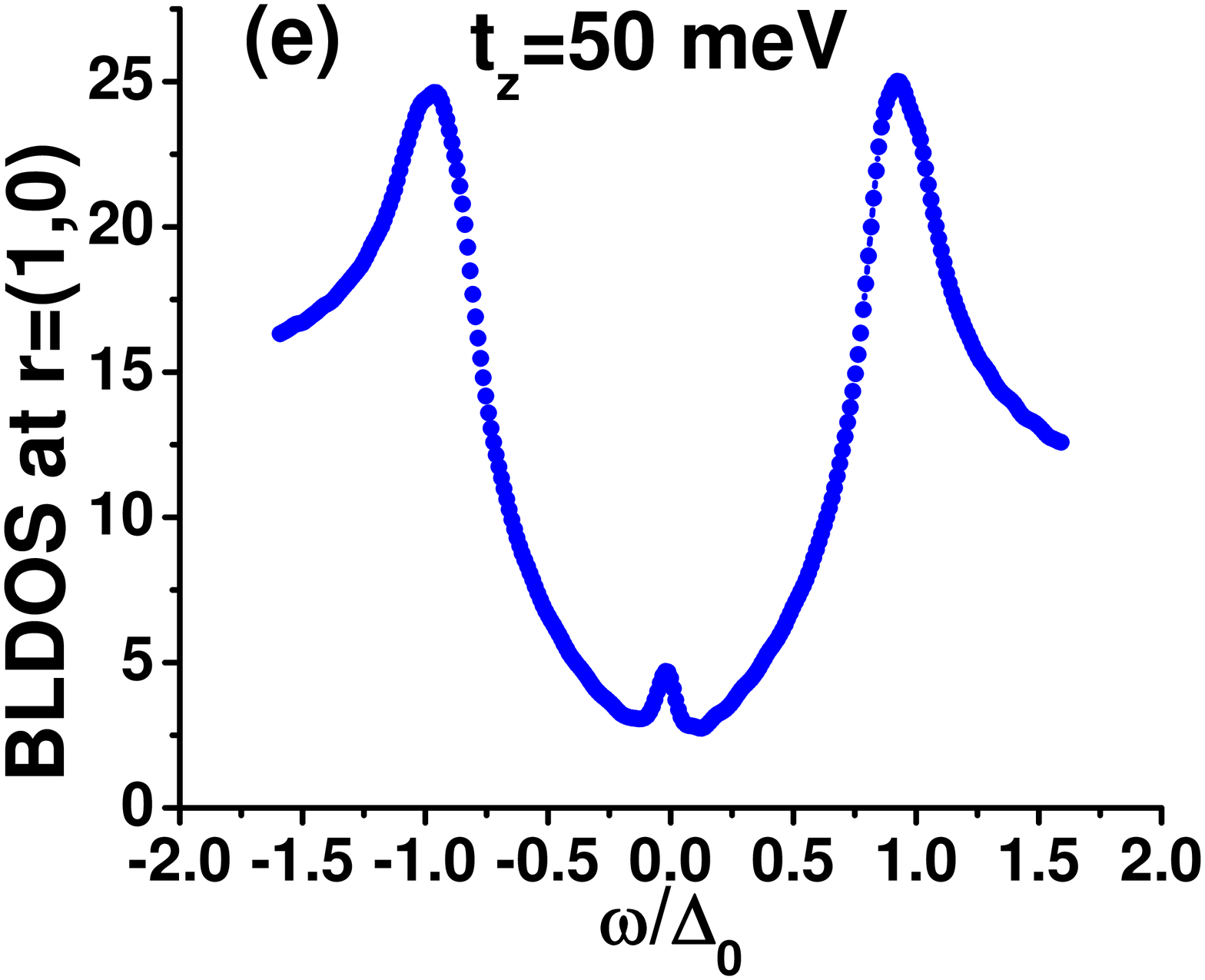}}
\rotatebox[origin=c]{0}{\includegraphics[angle=0, 
           height=1.3in]{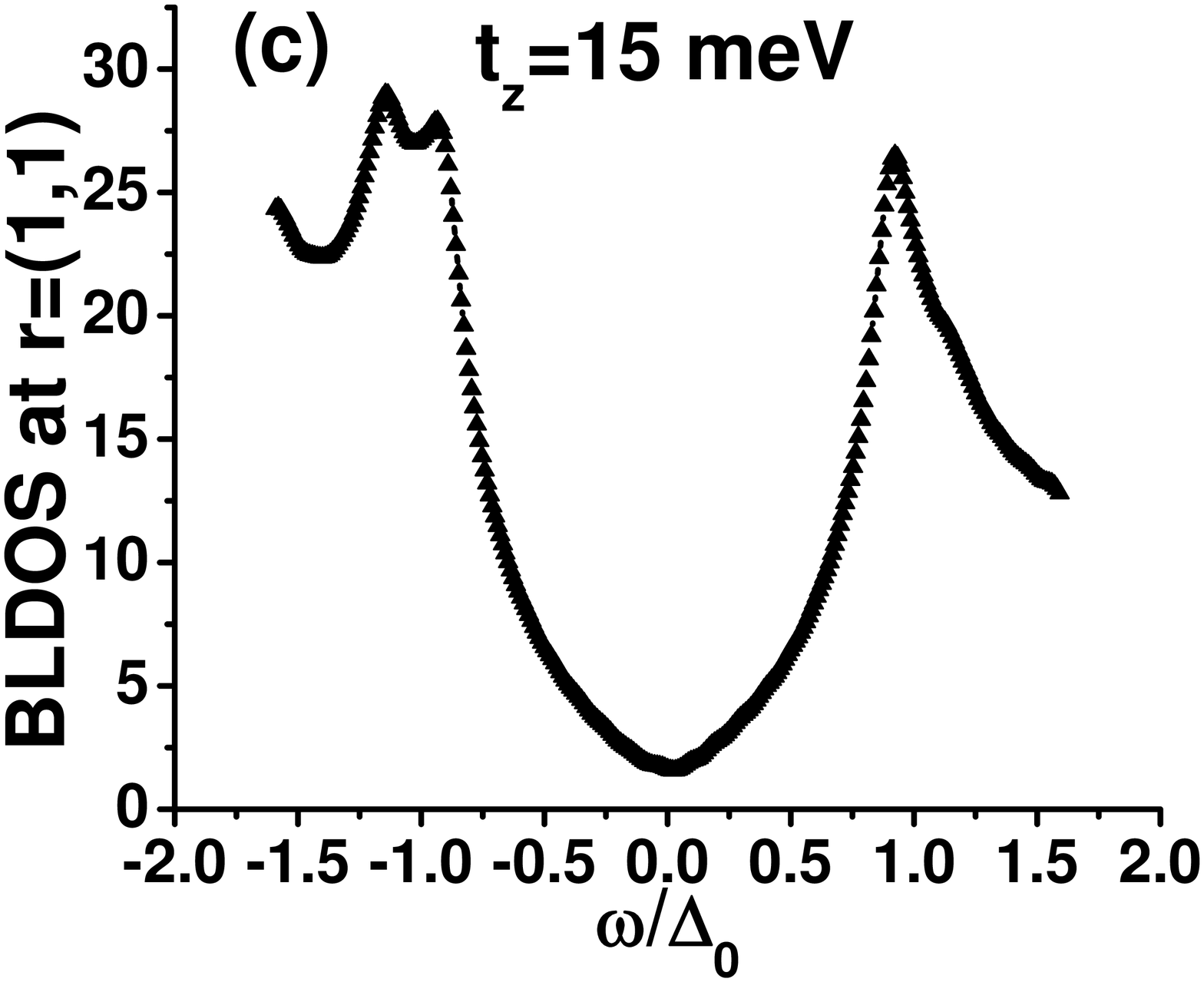}} 
\rotatebox[origin=c]{0}{\includegraphics[angle=0, 
           height=1.3in]{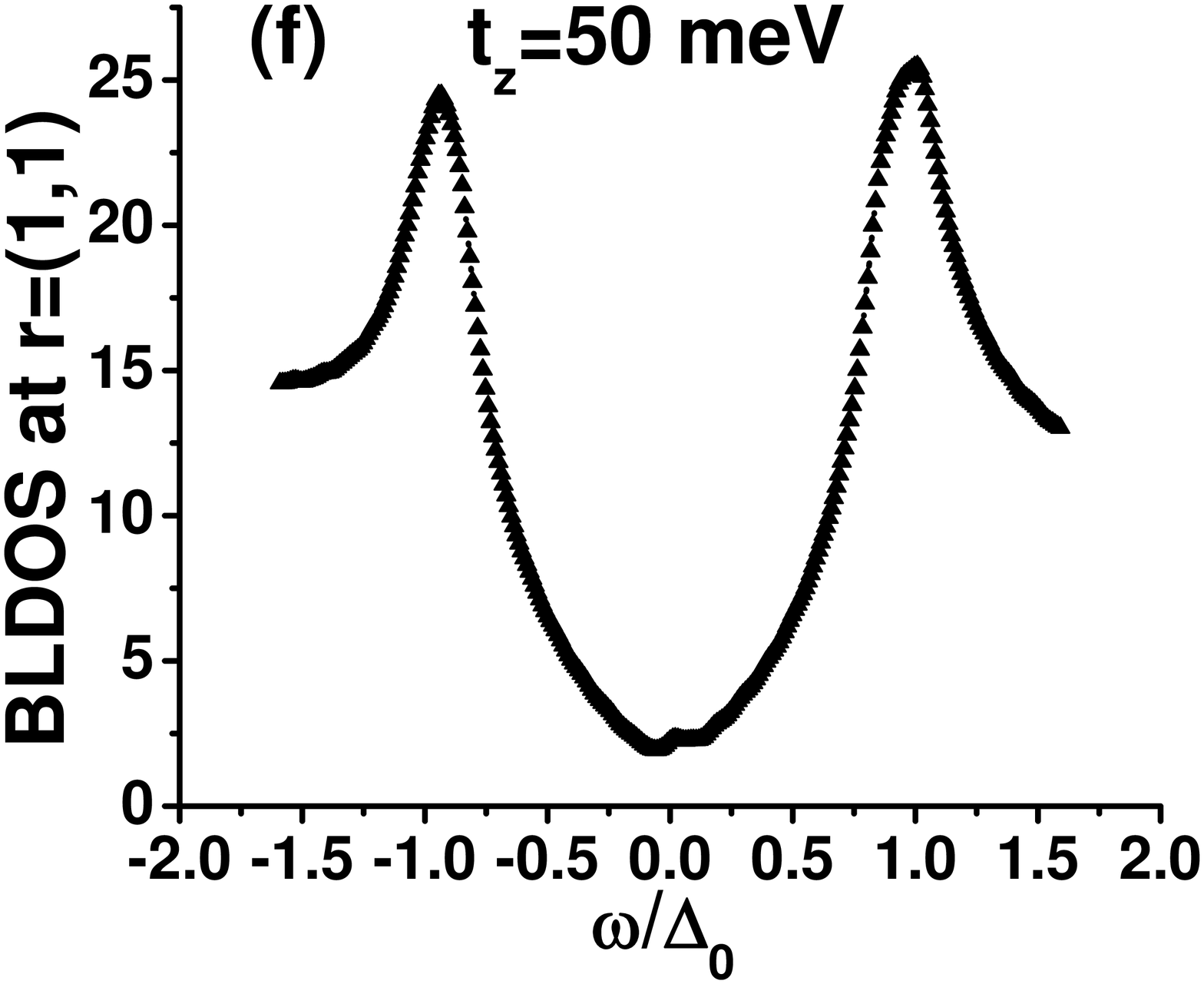}}              
\caption{(Color online) The predicted BLDOS $\rho_{\rm STM}({\bf r},\omega)$ at the sites (0,0), (1,0), and (1,1) with different tunneling strength $t_z$. The energy gap $\Delta_0=44$ meV and the doping $p\sim 15\%$.}           
\end{figure}

It is out of question that there exists a critical value $t_{zc}$ of the tunneling strength $t_z$, at which a resonant peak near the Fermi surface is first produced at (0,0). In Fig. 2, we show the critical value $t_{zc}$ as functions of energy gap $\Delta_0$ and doping $p$. If $p$ is fixed, $t_{zc}$ decreases with increasing $\Delta_0$. We note that when $p\sim 20\%$ $t_{zc}$ does not vary with  $\Delta_0$($>35$ meV). If $\Delta_0$ ($<$75 meV) is fixed, $t_{zc}$ decreases with increasing $p$. For the other band structures, including that with the nearest-neighbor and next-nearest-neighbor hopping, $t_{zc}$ has similar relations with $p$ and $\Delta_0$. 

\begin{figure}
\rotatebox[origin=c]{0}{\includegraphics[angle=0, 
           height=1.5in]{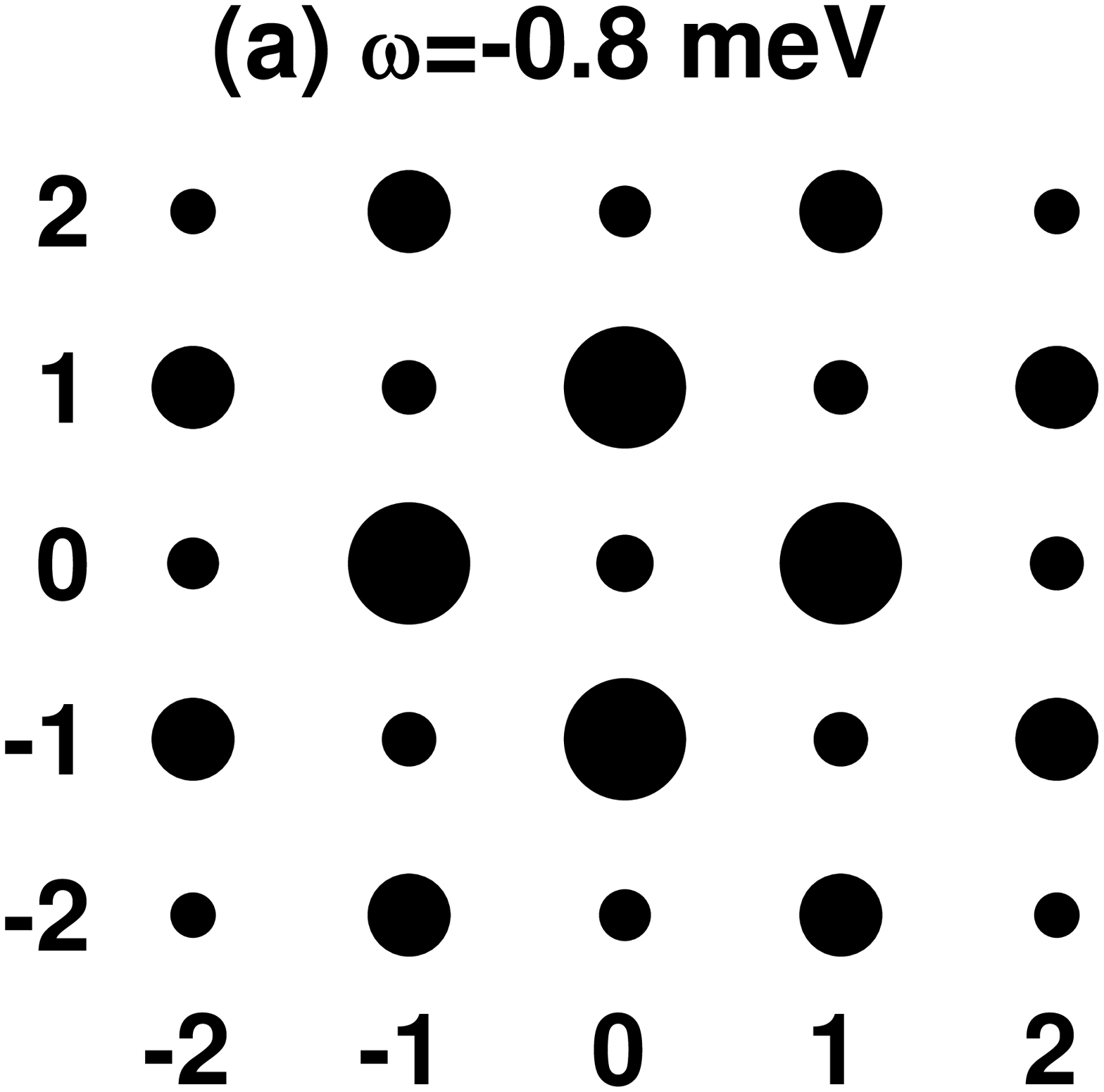}}
\rotatebox[origin=c]{0}{\includegraphics[angle=0, 
           height=1.5in]{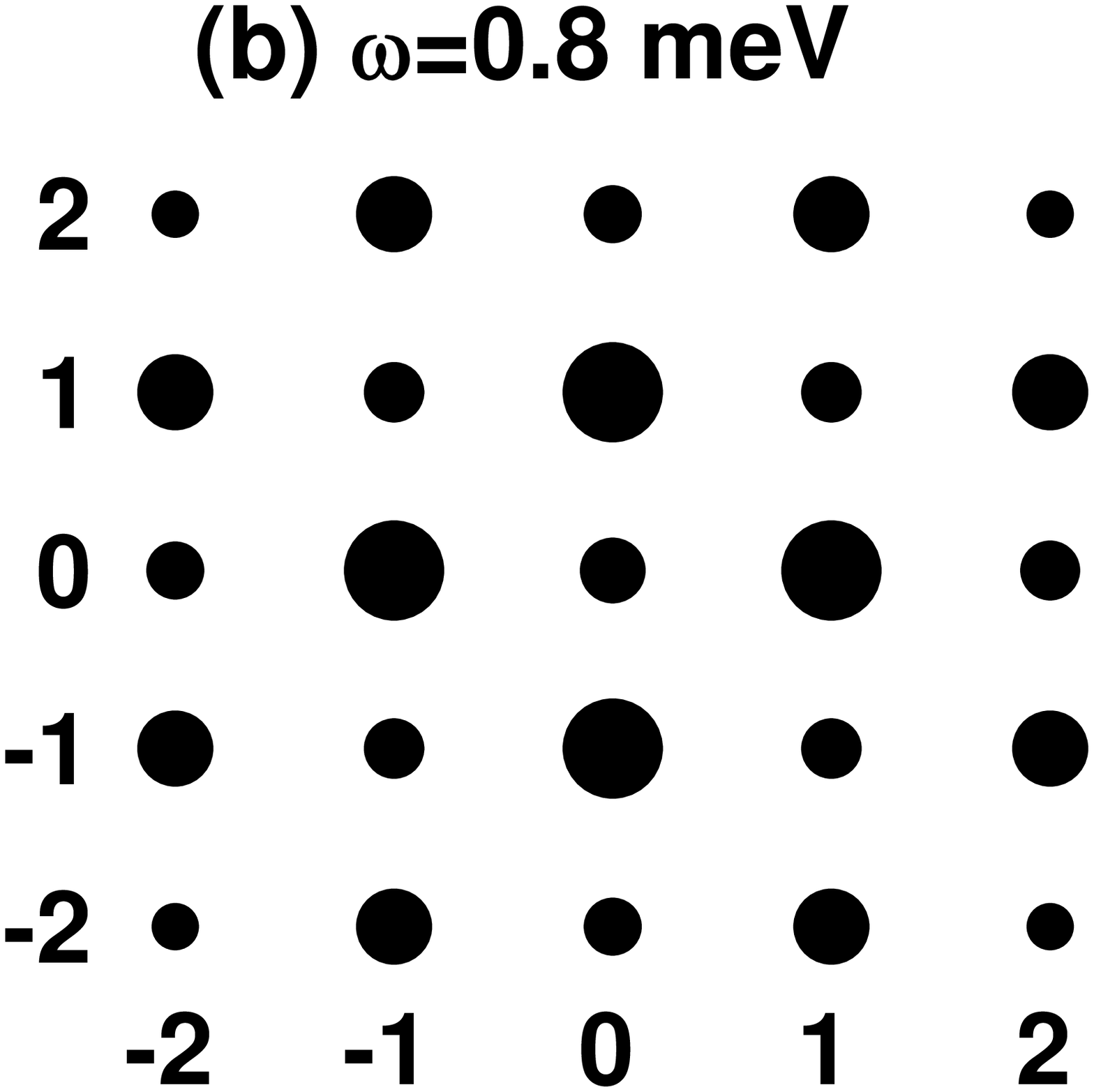}}
\caption{The patterns of $\rho_{\rm STM}({\bf r},\omega)$ near the site just above a Zn impurity at different energy $\omega$. The radius of a solid circle is proportional to the magnitude of $\rho_{\rm STM}({\bf r},\omega)$. The energy gap $\Delta_0=44$ meV, the doping $p\sim 15\%$, and the tunneling strength $t_z=50$ meV.}             
\end{figure}

Now we discuss the STM experiments on a bilayer cuprate, e.g. Bi2212. It is known that the surface of Bi2212 is a Bi-O layer and these Bi atoms locate just above Cu atoms in the first Cu-O layer. Therefore, the electrical current from the STM tip is blocked by Bi atom and is approximately proportional to the summation of the LDOS at four neraest neighboring sites of the Cu atom beneath the STM tip [12]. The blocking LDOS (BLDOS) $\rho_{\rm STM}({\bf r},\omega)\approx \sum_{\delta=\pm 1}[\rho({\bf r}+\delta {\bf i_x},\omega)+\rho({\bf r}+\delta {\bf i_y},\omega)]$, where ${\bf i_x}$ and ${\bf i_y}$ are the unit vectors along ${\bf x}$ and ${\bf y}$ directions, respectively. In Fig. 3, we plot the curves of $\rho_{\rm STM}({\bf r},\omega)$ at different sites. When $t_z=15$ meV,  $\rho_{\rm STM}({\bf r},\omega)$ has no resonant peak near the Fermi surface. The curves of $\rho_{\rm STM}({\bf r},\omega)$ and $\rho({\bf r},\omega)$ at the same site are similar below energy gap [see Fig. 2(a)-(c) and Fig. 3(a)-(c)].  When $t_z=50$ meV, $\rho_{\rm STM}({\bf r},\omega)$ has a resonant peak at (0,0) and (1,1) with $\omega=0.8$ meV and at (1,0) with $\omega=-0.8$ meV. Obviously, the highest resonant peak in $\rho_{\rm STM}({\bf r},\omega)$ is at (1,0) while that in $\rho({\bf r},\omega)$ is at (0,0). We note that the resonant peaks in $\rho_{\rm STM}({\bf r},\omega)$ and $\rho({\bf r},\omega)$ at the same site locate at opposite bias voltages (see Fig. 1(d)-(f) and Fig. 3(d)-(f)). In Fig. 4, we plot the images of $\rho_{\rm STM}({\bf r},\omega)$ near the site (0,0) with $t_z=50$ meV at $\omega=\pm 0.8$ meV. Obviously, $\rho_{\rm STM}({\bf r},\omega)$ oscillates with the distance from (0,0) and has a maximum at $(\pm 1,0)$ and $(0,\pm 1)$. These patterns also have a four-fold symmetry, but are opposite to those induced by a Zn impurity in the first Cu-O layer of Bi2212, where the resonant peak is highest at (0,0) [11]. 

It is known that the cuprate superconductors are usually inhomogeneous [18]. Different regions have different dopings and energy gaps. For Bi2212, the region with larger doping has a smaller gap [19]. From Fig. 2, we can see that the critical tunneling strength $t_{zc}$ to produce a resonant peak through the interlayer tunneling depends strongly on doping and energy gap. The smaller $t_{zc}$, the easier the resonant peak to be observed by the STM experiments. In order to see more clearly the tunneling-mediated resonant peak near the Fermi surface, we should choose the regions of Bi2212 with larger doping. This is consistent with the choice of samples in ARPES experiments [9,10]. In other regions with the tunneling strength smaller than the critical value, there is no resonant peak in the LDOS. From viewpoint of statistics, the numbers of the Zn impurities in the first and second Cu-O layers in Bi2212 are equal. Therefore, the number of the patterns of $\rho_{\rm STM}({\bf r},\omega)$ in Fig. 4 is much less than that induced by the Zn impurities in the first Cu-O layer in Bi2212 [20].     

In conclusion, we have solved analytically the Hamiltonian (1) by the Green's function method and calculated numerically the surface LDOS with different physical parameters. We have found the critical tunneling strength, above which the impurity resonant peak appears near the Fermi surface. The LDOS pattern is opposite to that produced by the Zn impurities in the first Cu-O layer of Bi2212 at the resonant bias voltage. The sign of the bias voltage at the resonant peak alternates with the distance from the site just above the Zn impurity in the second Cu-O layer. We hope that the tunneling-mediated impurity resonant peak and the corresponding LDOS pattern could be verified by the future STM experiments.

The authors would like to thank Professor S. H. Pan for useful discussions. This work was supported by the Texas Center for Superconductivity at the University of Houston 
and by the Robert A. Welch Foundation under the Grant no. E-1411.

%\end{multicols}

\end{document}